\documentclass[a4paper,11pt]{article}
\pdfoutput=1 % if your are submitting a pdflatex (i.e. if you have
\usepackage{jcappub} % for details on the use of the package, please
\usepackage[normalem]{ulem}      
\usepackage{bm}

\usepackage[T1]{fontenc} % if needed
\usepackage{bigints}
\usepackage{longtable}
\usepackage{ amssymb }

\newcommand{\beq}{\begin{equation}}
\newcommand{\eeq}{\end{equation}}
\usepackage{bm}
\newcommand{\de}{{\textrm d}}

\usepackage[T1]{fontenc}
\usepackage[utf8]{inputenc}
\usepackage{relsize}
\usepackage{soul}
\usepackage{calligra}
\usepackage{verbatim}
\usepackage{caption}
\usepackage{adjustbox}
\usepackage[titles]{tocloft} %
\cftsetindents{section}{0em}{2.5em}
\cftsetindents{subsection}{2.5em}{2.5em}

\usepackage{appendix}

\usepackage{graphicx,epsfig}
\usepackage[usenames,dvipsnames,table]{xcolor}
\usepackage{subfig}
\usepackage{float}
\usepackage{placeins}

\usepackage{tabularx}
\usepackage{makecell,multirow}
\usepackage{dcolumn}
\usepackage{enumitem}

\usepackage{amsmath,amsthm,amssymb}
\usepackage{slashed}
\usepackage{mathrsfs}
\usepackage{bm}
\usepackage{leftidx}
\usepackage{commath}
\usepackage{natbib}

\usepackage{url}
\usepackage{color}
\definecolor{darkblue}{rgb}{0.0,0.0,0.6}
\definecolor{red}{rgb}{0.9, 0,0}
\definecolor{navy}{rgb}{0.05, 0.05,0.8}
\hypersetup{
     colorlinks   = true,
     citecolor    = darkblue,
     linkcolor    = darkblue,
     urlcolor    = darkblue
}
\graphicspath{{./figures/}}

\newcommand{\tx}[1]{\ensuremath{\textnormal{#1}}}
\newcommand{\eqa}[1]{\begin{align}#1\end{align}}

\newcommand{\rp}{\right)}
\newcommand{\lp}{\left(}
\newcommand{\rb}{\right]}
\newcommand{\lb}{\left[}
\def\bit{\begin{itemize}}
\def\eit{\end{itemize}}
\def\ben{\begin{enumerate}}
\def\een{\end{enumerate}}

\newcommand{\Sec}[1]{Sec.~\ref{#1}}

\newcommand{\Fig}[1]{Fig.~\ref{#1}}

\newcommand{\units}[1]{\;\tx{#1}}
\newcommand{\eV}{\units{eV}}

\newcommand{\GeV}{\units{GeV}}

\newcommand{\cm}{\units{cm}}

\newcommand{\micron}{\mu \text{m}}

\newcommand{\kpc}{\units{kpc}}

\newcommand{\be}{\begin{equation}}
\newcommand{\ee}{\end{equation}}

\newcounter{questioncount}[section]

\newcounter{notecountRJ}[section]

\newcommand{\mdm}{m_\tx{DM}}

\def\lsim{\mathrel{\raise.3ex\hbox{$<$\kern-.75em\lower1ex\hbox{$\sim$}}}}
\def\gsim{\mathrel{\raise.3ex\hbox{$>$\kern-.75em\lower1ex\hbox{$\sim$}}}}

\begin{document}
\title{\boldmath Hunting Dark Matter Lines in the Infrared Background with the James Webb Space Telescope}

\author[a]{Ryan Janish}
\author[a,b]{Elena Pinetti} 
%\author{Elena Pinetti$^{1,2}$}
\emailAdd{rjanish@fnal.gov}
\emailAdd{epinetti@fnal.gov}
%\thanks{epinetti@fnal.gov, http://orcid.org/0000-0001-7070-0094}

\affiliation[a]{Fermi National Accelerator Laboratory, Theoretical Astrophysics Department, Batavia, Illinois, 60510, USA}
\affiliation[b]{University of Chicago, Kavli Institute for Cosmological Physics, Chicago, IL 60637, USA}

\abstract{Dark matter particles with a mass around 1$\eV$ can decay into near-infrared photons. 
Utilising available public blank sky observations from the NIRSpec IFU on the James Webb Space Telescope (JWST), we search for a narrow emission line due to decaying dark matter and derive leading constraints in the mass range 0.8-3$\eV$ on the decay rate to photons, and more specifically, on the axion-photon coupling for the case of axion-like particles. We exclude $\tau < 6.7\cdot 10^{26}$ s at $\mdm \simeq 0.9 \eV$ and, in the case of axions, $g_{a\gamma\gamma} > 9.4 \cdot 10^{-12} \GeV^{-1}$ for $m_a = 2.15 \eV$. 
Our results do not rely on dedicated observations, rather we use blank sky observations intended for sky subtraction, and thus our reach may be automatically strengthened as JWST continues to observe.
}

\maketitle
\flushbottom

\section{Introduction}
Revealing the nature of dark matter (DM) is one of the most 
important %ambitious 
goals of modern physics. Despite the overwhelming cosmological and astrophysical evidence of its existence, the microscopic properties of DM remain an enigma (see e.g. \cite{Boddy:2022knd} for a review). 
One enticing possibility is that DM is %made up unstable 
unstable and decays to Standard Model (SM) states 
on a cosmological timescale. 
Astronomical %Terrestrial and space-based 
telescopes are %represent 
ideal instruments to look for a signal from decaying DM in its %natural
cosmological environment, where we expect DM to be particularly abundant. 

%In this respect, 
DM candidates decaying into two particles including a photon are of particular interest
as they produce a nearly monochromatic emission line which 
can %since can more easily 
stand out from the astrophysical background. 
For instance, DM of mass $\mdm$
%particles could produce two photons, each with frequency of $\nu = m_a/4 \pi$.
may decay to two photons yielding a line of frequency $\nu =  \mdm/4 \pi$.
%As a result, 
The mass of the DM candidate 
then
determines the wavelength 
%where we expect to find a line emission. 
at which we expect to find an emission line. 
Various searches for these decay 
lines %products 
have been proposed over the years,
ranging from radio waves to X-rays \cite{Krnjaic:2023odw, Regis:2020fhw,Caputo:2020msf}. % more? 
In particular, DM %candidates 
with mass on the order of an electronvolt would produce photons in the infrared and optical bands. 
Previous 
searches in these bands have used spectroscopy %studies have focused on data 
from the Hubble Space Telescope, Spitzer and 
VLT %MUSE  note: muse is one instrument on the VLT telescope
~\cite{Todarello:2023hdk, Grin:2006aw, Carenza:2023qxh, Nakayama:2022jza, Caputo:2020msf}
and set the leading constraints on such DM decays for candidates with mass larger than roughly $2.6\eV$.
It is worth noting that infrared and optical detectors tend to focus on a relatively narrow band, compared to X-ray and gamma-ray telescopes,
%. For these reasons, 
so these constraints can typically improve over a restricted range of masses. 
In this work, and for the first time, we 
search %look 
for a DM-produced infrared line the in James Webb Space Telescope (JWST) 
spectroscopic data. %data. 
Launched in 2021, JWST is one of the most advanced infrared telescopes, 
sensitive to %focusing on 
wavelengths between $0.6\,\micron$ and $28.3\,\micron$, 
which corresponds to DM masses between $0.1\eV$ and $4.1\eV$ for a two-photon decay signal. 

We focus in this work on observations of the blank sky, rather than specific observations of DM-rich targets such as dwarf galaxies or the galactic center, as has been previously discussed~\cite{Bessho:2022yyu}. 
All blank sky observations look out through the Milky Way DM halo and are thus sensitive to a DM decay signal.
The strength of this signal is similar or even larger than that from dwarf galaxies.
In addition, the primary advantage of this approach is that blank sky observations accompany most astronomical observations regardless of the intended target, for the purposes of sky subtraction. a
We can thus immediately perform a DM search with the available public data by harvesting from it observations of the blank sky.
Further, our results may be readily updated and strengthened with future data, regardless of the targets chosen in future observing cycles and without need to obtain specific observing time.

Our DM search is motivated in particular by the possibility of discovering or constraining axion-like particle DM. 
Among the numerous theories beyond the Standard Model that predict compelling DM candidates, axions remain one of the most promising possibilities (see e.g. \cite{Adams:2022pbo} for a review). 
They are ultra-light 
pseudoscalar %bosonic 
particles which can interact with SM particles such as photons through a dimension-5 operator, and indeed generically decay directly to two photons. %electrons  
The concept of axions first emerged as a solution to the strong charged-parity (CP) problem in quantum chromodynamics~\cite{PhysRevD.16.1791, PhysRevLett.38.1440, PhysRevLett.40.279, PhysRevD.11.3583}. %A peculiar feature of the QCD axion is its coupling to photons via radiative decay emission. 
Numerous extensions of the SM 
contain %predict 
ultra-light pseudoscalar DM candidates that share the same photon coupling of the QCD axion but do not solve the strong CP problem, 
%. These candidates are known as axion-like particles 
known as axion-like particles (ALPs)~\cite{Adams:2022pbo},
and these represent 
%one practical example of decaying DM considered in this work.  %Axion-like particles (ALP) are among the most popular DM candidates. One notable example is the QCD axion, orignally proposed as a solution to the strong charged-parity problem.  
the primary example of decaying DM considered in this work.

The paper is structured as follows. Section \ref{sec:signal} describes the formalism for the computation of the DM signal. Section \ref{sec:JWST} introduces the James Webb Space Telescope and the observations that we utilise in our analysis. In Section \ref{sec:results}, we 
describe our analysis procedure and 
present the main results. 
Finally, in Section \ref{sec:conclusion} we draw the conclusions.
Throughout this work we use natural units, in which $\hbar = c = 1$, unless stated otherwise. 

\section{Dark Matter Signal}
\label{sec:signal}

We will search for an infrared emission line produced by DM.  
This is motivated in particular by ALPs, which can decay to two photons via the interaction 
\begin{equation}
\label{eq:Lalps}
    \mathcal{L} = -\dfrac{1}{4} g_{a \gamma \gamma} \, a \, F_{\mu \nu} \Tilde{F}_{\mu \nu} \; ,
\end{equation}
where $a$ represents the ALP field, $g_{a \gamma \gamma}$ is the ALP-photon coupling, and $F_{\mu \nu}$ and $\Tilde{F}_{\mu \nu}$ denote the electromagnetic field strength and its dual, respectively.
This decay has a rate \cite{Cadamuro:2011fd}
\begin{equation} 
\label{eq:alp-decay}
  \Gamma_{\gamma} = \frac{g_{a \gamma \gamma}^2 m_a^3}{ 64 \pi}
\end{equation}
for an ALP of mass $m_a$. 
Our search, however, is sensitive to any DM candidate which produces a suitably narrow emission line. 
We can parameterize the strength of such emission by the total luminosity produced per DM mass, which we call the \emph{emission rate} and denote $\Gamma_\gamma$. 
For decaying DM, $\Gamma_\gamma$ is simply the decay rate to photons. 
The resulting differential energy flux of photons observed by a particular instrument is 
\begin{align}
\label{eq:general-flux}
  \Phi_\tx{DM} = \frac{\de \phi}{\de \nu \de \Omega} &= 
  \frac{\Gamma_\gamma}{4\pi} 
  \left( \frac{\de f}{\de \nu} * W \right) D  \; ,
\end{align}
where $D$ denotes the so-called D-factor, i.e.~the integral of the DM mass density $\rho$ along the line-of-sight. The emission spectrum $\de f / \de \nu$ specifies the fraction of the total luminosity that is emitted within a frequency interval $\de \nu$.  
For simplicity we have assumed that $\de f/ \de \nu$ is uniform in space. 
The spectrum $\de f / \de \nu$ is convolved with the instrumental response function $W$ to produce the observed spectrum.   

In this work we consider the decay of DM particles in the Milky Way halo.
The DM density is isotropic with respect to the galactic center, so the D-factor is computed for a given line-of-sight as 
\eqa{
\label{eq:Dfactor}
   D\lp \theta \rp = \int\displaylimits_0^\infty \de s \; \rho\lp r\lb s, \theta \rb \rp \; ,
}
where $\theta$ is the angle between the line-of-sight and the galactic center, r is the galactocentric distance, and $s$ is the distance along the line-of-sight.  
These quantities are related by $r^2 = s^2 + R_\odot^2 - 2 s R_\odot \cos \theta$, where we have taken the galactocentric distance of the Sun to be $R_\odot = 8.1 \kpc$.
For a line-of-sight with galactic latitude $b$ and longitude $\ell$, the angle $\theta$ is given by $\cos \theta = \cos \ell \cos b$.
We assume a standard NFW DM profile~\cite{Navarro:1995iw,Cirelli:2010xx} 
\begin{equation}
    \rho(r) = \dfrac{\rho_s}{\left(\dfrac{r}{r_s} \right) \left(1 + \dfrac{r}{r_s} \right)} \; , 
\end{equation}
where we take $r_s = 24 \kpc$ and $\rho_s = 0.18 \GeV/\cm^3$~\cite{Cirelli:2020bpc}.
In this work we do not consider lines-of-sight that pass near the galactic center, and so we are insensitive to the exact behaviour of $\rho$ as $r \rightarrow 0$. 

We consider the case of monochromatic emission in the DM rest frame, so that $\de f / \de E$ is set only by Doppler broadening.
The fractional width of the line is roughly $10^{-3}$, given by the velocity dispersion of the Milky Way. 
This is the case for decay to two photons, which results in a line at frequency $\nu_0 = \mdm/4\pi$. 
The Doppler width is comparable to the fractional resolution of the medium resolution gratings of JWST NIRSpec used in this work, which have $\Delta \lambda / \lambda \approx 10^{-3}$~\cite{NIRSpec-Dispersers-Filters} (more details in Sec. \ref{sec:JWST}), and so we include the Doppler line shape $\de f/ \de E$ in this analysis.  
Furthermore, other JWST instruments and configurations have even higher resolution, so future analyses will require this. 
The Doppler line shape in the galactic frame is given by  
\eqa{
    \frac{\de f}{\de \nu} = \frac{\nu}{2 \nu_0^2} 
    \int\displaylimits_{v_\tx{min}(\nu)}^\infty 
    \frac{f\lp v \rp}{v} \;  \de v , \quad\quad 
      v_\tx{min}(\nu) = \frac{|\nu - \nu_0|}{\nu_0}
}
for a line centered on $\nu_0$ and an isotropic DM speed distribution $f(v)$.   
We adopt a uniform, isotropic Maxwellian speed distribution, 
\eqa{
\label{eq:maxwell_f}
  f \lp v \rp = \frac{4 \pi v^2}{\lp 2 \pi \sigma_v^2 \rp^{3/2}} 
  \; e^{-v^2/2 \sigma_v^2} 
}
with dispersion $\sigma_v = 160$ km/s ~\cite{Evans:2018bqy}. 
This is a good approximation for lines-of-sight that do not pass near the galactic center, as we consider here. 
We neglect to truncate this distribution at the escape velocity, $v_\tx{esc} = 510$ km/s $\approx 3.1 \sigma_v$~\cite{Drukier:1986tm, Evans:2018bqy}, as the tails of this distribution are irrelevant in our analysis. 
% Equation~\eqref{eq:maxwell_f} yields a line profile  
% \eqa{
%     \frac{\de f}{\de \nu} = 
%   \frac{\nu}{\lp 2 \pi \nu_0^2 \sigma_v^2 \rp^{1/2} \nu_0} 
%   e^{- \lp \nu - \nu_0 \rp^2 / 2 \nu_0^2 \sigma_v^2}
% }
Equation~\eqref{eq:maxwell_f} yields an approximately Gaussian line profile, which expressed as a function of wavelength is  
\eqa{
    \frac{\de f}{\de \nu} \approx 
  \frac{\lambda_0^2}{\lp 2 \pi w^2 \rp^{1/2}} 
   \; e^{- \lp \lambda - \lambda_0 \rp^2 / 2 w^2} ,
}
where $\lambda_0 = 1/\nu_0$ and $w = \lambda_0 \sigma_v$.
We have made use of the fact that $\lambda_0 \gg w$ to simplify $\lambda^{-1} \approx \lambda_0^{-1} - (\lambda - \lambda_0)/\lambda_0^2$.
For the instrumental response we take a Gaussian in wavelength~\cite{2022A&A...661A..80J, NIRSpec-Dispersers-Filters}, 
\eqa{
    W(\lambda) = 
    \frac{1}{\lp 2 \pi \sigma_\lambda^2 \rp^{1/2}}
    \; e^{-\lambda^2/2 \sigma_\lambda^2} \; ,
}
where $\sigma_\lambda$ is the instrumental dispersion, related to the FWHM resolution $\Delta \lambda$ as  $\sigma_\lambda = \Delta \lambda / 2 \sqrt{2 \ln{2}}$. 
The observed line profile is thus a Gaussian with dispersion $\sigma^2 = \sigma_\lambda^2 + w^2$.

\begin{figure*}[t]
\centering
\includegraphics[width=\columnwidth]{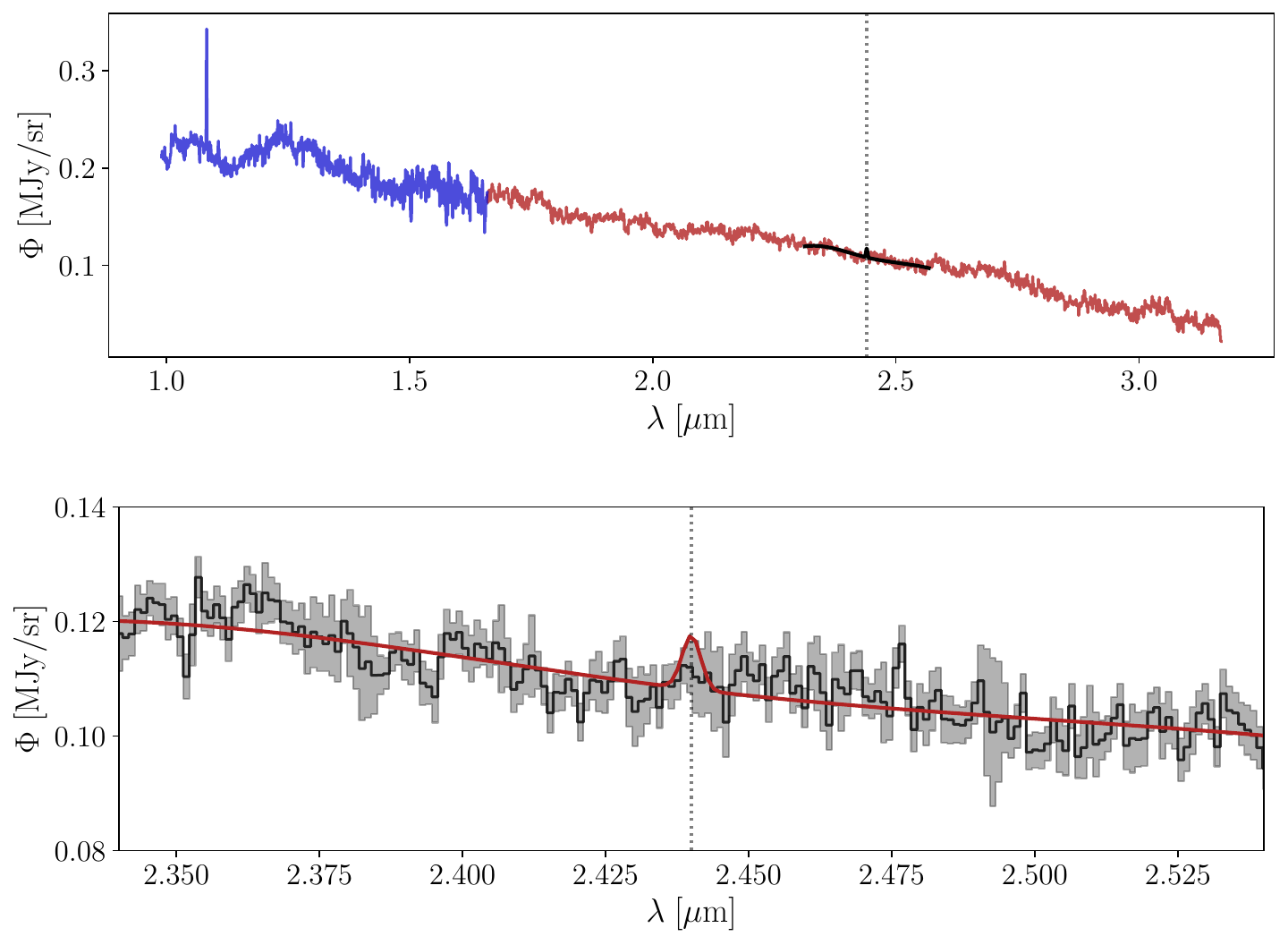}
\caption{{\bf Top:} The sky spectrum that we used in this work, collected in association with observations of the target GN-z11 by JWST NIRSpec using filter/grating combinations of F100LP/G140M (blue) and F170LP/G235M (red). 
See text of \Sec{sec:JWST} for details.
The solid black line illustrates an example of a continuum fit and DM line at $\lambda = 2.44 \, \micron$, which is also indicated by the dotted gray vertical line.
The width of the black line extends over the continuum modelling sub-region. 
See \Sec{sec:results} for details.
{\bf Bottom:} The same spectrum and example of DM line, now shown over only a fraction of the modelling region and with the observed spectrum in black, its errors in gray, and the DM model in red. 
The strength of the depicted DM line is allowed at 2$\sigma$, i.e~our analysis would rule out any stronger signal.
This DM example corresponds to an ALP of $m_a = 1 \, \eV$ and $g_{a\gamma\gamma} = 1.1 \cdot 10^{-11} \, \GeV^{-1}$
}
\label{fig:full-spectra}
\end{figure*}

\section{James Webb Space Telescope Observations} 
\label{sec:JWST}

Launched on 25 December 2021, the James Webb Space Telescope \cite{Gardner:2006ky} is a high-sensitivity and high-resolution infrared telescope. It covers the wavelength range 0.6–28.3 $\micron$, corresponding to the energy range 2-44 meV. It carries four scientific instruments capable of a variety of modes of spectrography: Near-Infrared Camera (NIRCam, 0.6–5 $\micron$), Near-Infrared Spectrograph (NIRSpec, 0.6–5 $\micron$), Mid-Infrared Instrument (MIRI, 4.9–28.5 $\micron$) and Near-Infrared Imager and Slitless Spectrograph (NIRISS, 0.6–5 $\micron$). 
For a blank-sky DM search, we require high spectral resolution covering a large field-of-view. 
Spatial resolution is irrelevant, save for the removal of contaminating sources.
Grism spectroscopy is unsuitable as each point in the field is an equally bright source of DM signal which produces overlapping spectra on the detector, however both integral field unit (IFU) spectroscopy and multi-object spectroscopy (MOS) work well here. 
IFU is a technique that combines spectroscopy and imaging. 
In the course of an IFU observation, the apparatus records a wide field-of-view image in which each pixel simultaneously collects a spectrum. 
With MOS, many individual spectra are collected from a variety of configurable sky locations. 
MIRI and NIRSpec both have IFU modes, while NIRSpec also has a MOS mode, which are all well-suited for a blank-sky DM search.  

In this work we focus on a select pair of NIRSpec IFU observations, and leave the incorporation of all available IFU and MOS data for future work.
We chose to start with NIRSpec over MIRI as the background is considerably smaller~\cite{JWST-background} and the near-IR spectral range has a larger relative ALP decay rate~\eqref{eq:alp-decay}.
The NIRSpec IFU~\cite{Rauscher:2007ta, 2022A&A...661A..80J, 2022A&A...661A..82B} covers a field of view of 3" x 3" in spatial pixels of size 0.1" x 0.1", with a spectral resolving power of either $\lambda/\Delta \lambda \sim 100, 1000, 2700$, depending on the grating used. 
The DM sensitivity is a strong function of the resolution, and it is desirable for the resolution to at least meet the Doppler width of a DM line, $\lambda/\Delta \lambda \gtrsim \sigma_v^{-1} \sim 1000$.
We obtained public data from JWST DDT Program 4426 in Cycle 1%(PI: Roberto Maiolino)
, which has previously been used in Refs.~\cite{Maiolino:2023wwm, 2023arXiv230609142S}. 
These are NIRSpec IFU observations of the high-redshift galaxy GN-z11 observed on the 22nd and 23rd of May 2023, using two medium-resolution ($\lambda/\Delta \lambda \sim 1000$) filter/grating combinations~\cite{NIRSpec-Dispersers-Filters}:
\begin{itemize}
    \item[--] F100LP/G140M covering 0.97 - 1.9 $\micron$ with resolution $\Delta \lambda_\tx{G140M} \approx 1.4 \cdot 10^{-3} \, \micron$ 
    \item[--] F170LP/G235M covering 1.7 - 3.2 $\micron$ with resolution $\Delta \lambda_\tx{G235M} \approx 2.3 \cdot 10^{-3} \, \micron$ \; .
\end{itemize}
This set was chosen as an ideal starting point, having a large integration time, sufficient spectral resolution, and being a compact source so that much of the field is blank sky.  
The target is located at galactic latitude $b=54.8^\circ$ and longitude $l=126^\circ$,\footnote{Note that the absorption in this target location is negligible \cite{Schlegel:1997yv, extinction}} which gives a line-of-sight away from the galactic center with D-factor $D \approx 7.5 \GeV \kpc / \cm^3 \approx 1.5 \cdot 10^8 M_\odot/\kpc^2$. 
We use the extracted one-dimensional spectra available for these observations on MAST under obsID 139257277 and 139257240, in which the pixels corresponding to the target object and the background sky have been isolated and separately summed. 
We use the F100LP/G140M spectrum between 0.99 - 1.66 $\micron$ and F170LP/G235M between 1.66 - 3.2 $\micron$, which have integration times of $1167$ s and $1897$ s, respectively. 
These ranges do not cover the full spectral window of NIRSpec, however we leave searching in the rest of the spectral window for future work. 
This data is provided in the rest frame of the solar barycenter, and so we first transform it into a galactocentric frame, matching the expressions in~\Sec{sec:signal}.
The resulting sky spectrum is shown in \Fig{fig:full-spectra} and is roughly as expected from studies of the JWST background~\cite{2012ApJ...754...53K, 2023PASP..135d8002R,JWST-background}, being dominated by zodiacal light.  
The data described here may be obtained from~\url{http://dx.doi.org/10.17909/j5c2-c258}.

\begin{figure*}[t]
\includegraphics[width=0.95\columnwidth]{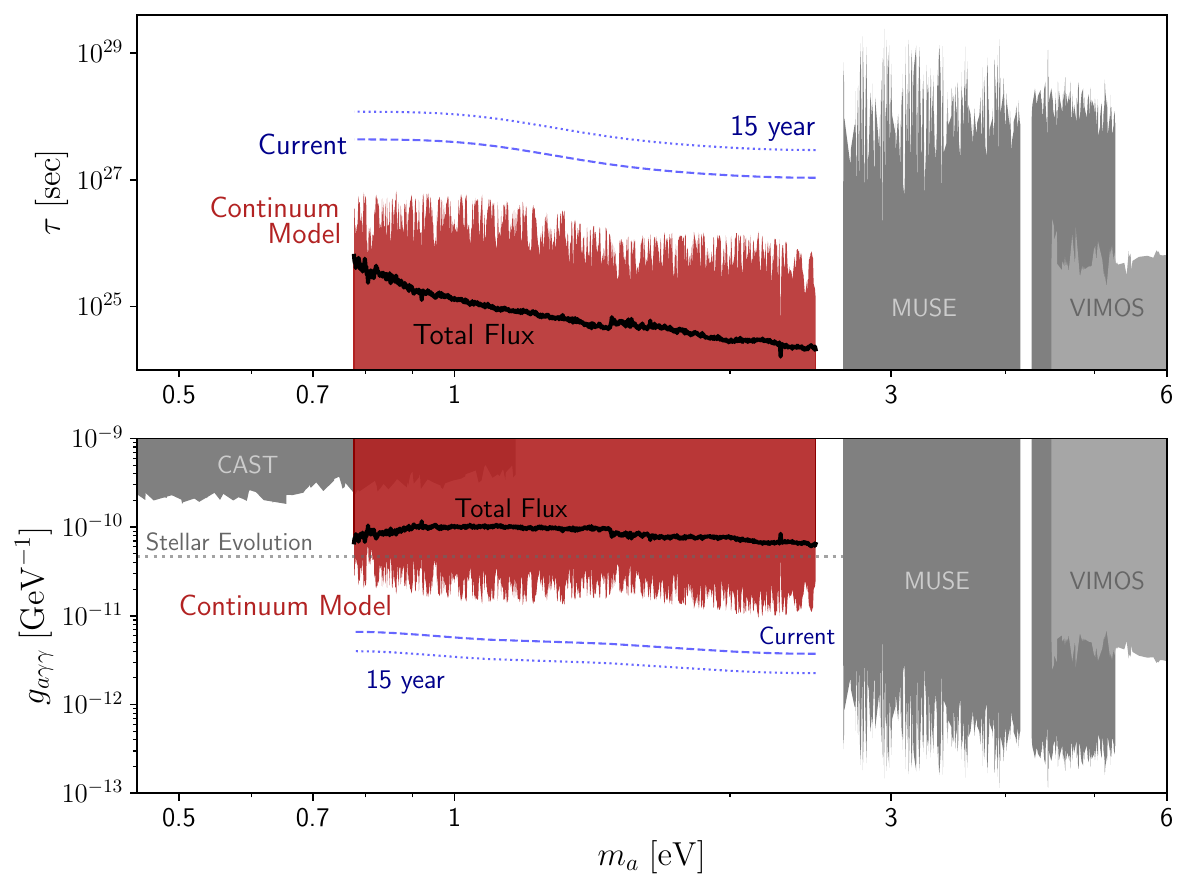}
\caption{
{\bf (Top)}
Our results, shown as 2$\sigma$ power-constrained limits on $\tau = 1/\Gamma_\gamma$ as a function of the DM mass $\mdm$, where $\Gamma_\gamma$ is the DM line luminosity per mass and 
$\tau$ is equivalent to the DM to two photon decay rate. 
{\bf (Bottom)} 
Our 2$\sigma$ power-constrained limits on the ALP-photon coupling $g_{a\gamma\gamma}$ as a function of ALP mass $m_a$, assuming a single ALP species comprises all of galactic DM.
Constraints from the total flux are shown in black and constraints using a smooth continuum model in red. 
In blue is shown the estimated reach after including all JWST data currently observed with NIRSpec using the filters and gratings considered here (dashed), as well as after fifteen years of operating time (dotted). 
See text for details. 
The shaded grey regions are already constrained by indirect searches for DM decay by MUSE~\cite{Todarello:2023hdk} and VIMOS~\cite{Grin:2006aw}, and helioscope searches for ALPs by CAST~\cite{CAST:2007jps, CAST:2017uph}. 
The region above the dotted grey line is disfavoured by considering the effect of ALP emission on stellar evolution~\cite{Ayala:2014pea, Dolan:2022kul}, however there is considerable complexity in setting such bounds~\cite{2023ApJ...943...95S}.
Existing constraints are taken from the compilation in Refs.~\cite{AxionLimits, Caputo:2021eaa}. 
}
\label{fig:limits}
\end{figure*}

\section{Analysis and Results}
\label{sec:results}
Here we 
% derive constraints on the 
search for and constrain 
DM line emission, thereby placing bounds on the DM decay lifetime and on the coupling $g_{a\gamma\gamma}$ in the case of ALPs. 
% We do not attempt a discovery analysis, i.e.~a search for an emission line appearing in all lines-of-sight with amplitude proportional to the D-factor of Equation~\eqref{eq:Dfactor}. 
% We leave that to future work, as it necessitates modelling all known astrophysical lines within our spectral range. 
We derive two epistemologically distinct bounds.
%on the line emission, one more conservative and one more aggressive. 
The first we refer to as the \emph{total flux} limit, which is a conservative bound roughly requiring that the DM flux~\eqref{eq:general-flux} does not exceed the total observed flux.  
This makes no attempt at modelling the background and is in that sense conservative.
The second bound constrains the presence of the DM flux~\eqref{eq:general-flux} on top of a generic, smooth continuum.  
This we call the \emph{continuum model} constraint.
It allows the possibility of a DM line discovery, and in the absence of a detection it provides a more comprehensive and generically stronger bound as it makes use of the expected narrow width of the DM signal.   
In essence, this bound limits the DM flux to not exceed the size of the fluctuations in the observed spectrum. 
Note that this limit grows stronger when more data is added to the analysis, whereas the total flux limit does not. 
The continuum model limit is robust, as the key assumption on which it rests is that the observed spectrum is dominated by continuum emission which varies on a scale much larger than the DM linewidth, which is expected from prior studies of JWST background~\cite{2012ApJ...754...53K, 2023PASP..135d8002R}.
Nonetheless we find it informative to provide both constraints.  

\paragraph{Total flux constraints} We set total flux bounds on the emission rate $\Gamma_\gamma$ by defining the following test statistic~\cite{Cowan:2010js, Cirelli:2020bpc, Roach:2022lgo}:
\beq
\chi_>^2 = \sum_i \left(\frac{{\rm max}[\Phi_{{\rm DM},i}(\Gamma_\gamma, m_{\rm DM})-\phi_i,0]}{\sigma_i}\right)^2,
\label{eq:chi2}
\eeq
where $\Phi_{{\rm DM},i}$ denotes the predicted photon flux from DM, $\phi_i$ is the observed flux and $\sigma_i$ its uncertainty, and the sum runs over all observed wavelengths $\lambda_i$.
We fix the DM mass $m_\tx{DM}$, making $\chi^2_>$ a function of $\Gamma_\gamma$ alone.
With Gaussian errors, this statistic is equivalent to a log-likelihood for the model $\Phi_{\rm DM}$ described in \Sec{sec:signal} with the model-constraint $\Gamma_\gamma > 0$, and it follows a $\chi^2$-distribution with one degree-of-freedom.
We impose a $2\sigma$ constraint on $\Gamma_\gamma$ such that  $\chi_>^2=4$, and then scan over all $m_\tx{DM}$ within our spectral range. 
% This scheme allows us to not include any astrophysical background, and it is therefore useful to derive  conservative limits. 
Our results are shown in \Fig{fig:limits}.

\paragraph{Continuum model constraints}  
We search for a DM line by defining a test statistic $\chi^2$, following Refs.~\cite{Cowan:2010js, Cirelli:2020bpc, Roach:2022lgo}.
We include a smooth continuum model $P$ with parameters $\beta_j$ and compute at fixed $\mdm$
\eqa{
\label{eq:continuum-chi2}
    \chi^2(\Gamma_\gamma, \beta_j) = \sum_i \lp 
    \frac{\Phi_{\tx{DM}, i}(\Gamma_\gamma, m_\tx{DM}) + P_i(\beta_j) - \phi_i}{\sigma_i} \rp^2, 
}
determining the best-fit $\hat{\Gamma}_\gamma > 0$ and $\hat{\beta_j}$ by minimising $\chi^2(\Gamma_\gamma, \beta_j)$. 
We compute the $N\sigma$ significance of a detected DM line via 
\eqa{
\label{eq:delta-chi2-detection}
  \Delta \chi^2 = \chi^2 (\hat{\Gamma}_\gamma, \hat{\beta}_j) 
     - \min_{\beta_j} \lb \chi^2 (0, \hat{\beta_j}) \rb = N^2\; .
}
Scanning over the DM mass in wavelength steps equal to the FWHM of the DM line, we find three lines with a local detection significance exceeding $5\sigma$, located at $1.083 \, \micron$, $1.405 \, \micron$, and $2.583 \, \micron$.
The first of these is visible in \Fig{fig:full-spectra}, and corresponds to a known Helium line~\cite{2015ApJ...808..124S, 2018ApJ...855L..11O}.
The latter two are much weaker, with local significance of $5.0\sigma$ and $5.1\sigma$ respectively and a global significance over $906$ distinct mass tests of $3.5\sigma$ and $3.6\sigma$ respectively.  
We do not consider these strong detections and do not consider here the possible SM backgrounds for these two lines, but instead place robust upper bounds on the DM emission.  

To set upper bounds we consider the test statistic $\Delta \chi^2$ ~\cite{Cowan:2010js, Cirelli:2020bpc, Roach:2022lgo} defined as
\eqa{
\label{eq:delta-chi2}
  \Delta \chi^2(\Gamma_\gamma) = \max_{\beta_j} \lb \chi^2 (\Gamma_\gamma, \beta_j) \rb 
     - \chi^2 (\hat{\Gamma}_\gamma, \hat{\beta_j}) \; .
}
As argued in the case of the total flux bound, for the model $\Phi_\tx{DM}$ in \Sec{sec:signal} with the model-constraint $\Gamma_\gamma > 0$, the statistic $\Delta \chi^2$ defined here is equivalent to a log-likelihood ratio between the models $\Phi_\tx{DM} + P$ with vanishing or positive $\Gamma_\gamma$, respectively, with nuisance parameters $\beta_j$ and assuming the errors are Gaussian distributed. Thus,
$\Delta \chi^2$ follows a $\chi^2$-distribution with one degree-of-freedom. 
We set a 2$\sigma$ upper-limit on $\Gamma_\gamma$ by taking $\Delta \chi^2 = 4$, scanning over all $\mdm$ in our spectral range. 
%\RJ{In https://arxiv.org/pdf/2207.04572.pdf they use 2.71 instead of 4. Why?}
Note that these limits are robust in the presence of an astrophysical emission line which mimics the DM signal, as the best-fit $\hat{\Gamma}_\gamma$ will model this line and then $\Delta \chi^2$ provides a constraint on the DM emission in addition to this contribution. 
% The drawback of these constrains is that they are overly-conservative and will obscure a true DM signal.
% A discovery search plus stronger limits may be derived by including a model of astrophysical lines, however we leave that to future work and focus here on setting robust limits.

We further power-constraint our bounds, as in Ref.~\cite{Roach:2022lgo} and following the procedure of Ref.~\cite{Cowan:2011an}, in order to avoid spurious exclusions due to downward fluctuations in the data. 
This also serves to make our limits robust against the presence of background astrophysical absorption lines. 
For each DM mass we generate simulated, Gaussian distributed data with mean given by the best-fit continuum model $P$ with no DM component, $\Gamma_\gamma = 0$, and derive from this data an upper bound on $\Gamma_\gamma$ by the above procedure.  
Repeating this yields a distribution of upper bounds $f(\Gamma_\gamma)$ and we set the power-constrained threshold $\Gamma_\gamma^*$ by%~\cite{Cowan:2011an}
\eqa{
    \int_0^{\Gamma_\gamma^*} \de \Gamma_\gamma \, f(\Gamma_\gamma) = 0.1587.
}
That is, the probability to derive a bound less than $\Gamma_\gamma^*$ is roughly 16\%.  
Our final upper bound is given by the maximum of $\Gamma_\gamma^*$ and the solution to $\Delta \chi^2(\Gamma_\gamma) = 4$.

We use a cubic-spline local background model $P(\beta_j)$ where $\beta_j$ are anchor-points at a fixed set of wavelengths $\lambda_j$: $P(\lambda_j) = \beta_j$. 
This is only a local model meant to capture the smooth general behaviour of the spectrum, so we do not attempt to fit the full spectrum using many $\beta_j$. 
Instead for each $\mdm$ we apply the above analysis in a sub-region of wavelengths centered on the decay wavelength $\lambda_\tx{DM} = 4\pi / \mdm$ with a width 150 times the FWHM of the DM model line. 
We include $n_j = 5$ anchor-points for $P$ in this region, equally spaced in $\lambda$ and including one on either endpoint of the analysis interval. 
Our results are insensitive to varying the number of anchors in $P$, as long as we take $n_j \geq 3$ and do not include so many as to allow $P$ to vary on the scale of the DM linewidth.  
This procedure is illustrated in \Fig{fig:full-spectra}, and our results are presented in \Fig{fig:limits}.

The error estimate included with the data is smaller than the typical size of fluctuations present in the spectrum. 
We re-estimate the error from the data by computing a clipped standard deviation of the residuals of a continuum-only fit within each analysis interval.
This increases the error by a factor between 2-3, which we apply as a uniform re-scaling of the provided error estimates. 
The clipping iteratively masks regions with width larger than the instrumental FWHM in which all the data deviate in concert from the continuum by more than three times the error, which avoids biasing this estimate due to line regions in which the continuum model is a poor fit. 
These regions are then excluded from the modelling, however in practice we find only a few such regions and their influence on the final result is negligible.

\paragraph{Projections}
The analysis presented here employs data with an integration time of roughly $2000$ s per data point. 
The continuum model reach will strengthen as more data is added, and a key virtue of a blank sky search is that a large fraction off all JWST observations will contain blank sky.
Thus our results may be immediately improved by including all currently available data, and then progressively updating over the lifetime of the instrument. 
We include in \Fig{fig:limits} an estimate of the reach of such efforts, given by scaling the results of this work to a new total combined integration time according to 
\eqa{
    \Gamma_\gamma &\propto t_\tx{int}^{1/2} \\
    g_{a\gamma\gamma} &\propto t_\tx{int}^{1/4},
}
i.e.~we assume that the modelling of many sky spectra will correspond roughly to a decrease in the effective measured error of the infrared background scaling as the square-root of amount of data used. 
After scaling, we smooth the projection in $\log g_{a\gamma\gamma}$ to remove the structure which is idiosyncratic to the particular pattern of fluctuations present in the data considered here. 
In \Fig{fig:limits} we present estimates for an analysis of all currently collected data, which we label \emph{current}, as well as an estimate for the reach of all data collected over the lifetime of JWST, which we take to be 15 years.  
We estimate the total integration time to be 1\% of the mission time, conservatively accounting for overheads and the time spent observing with other instruments and configurations. 
We refrain from extending our projection to cover the full spectral window of NIRSpec, corresponding to masses of 0.5$\eV$ to 4.1$\eV$, nor to that of MIRI, 0.1$\eV$ to 0.5$\eV$.
However it is expected that searches using additional data will have competitive reach across the NIRSpec band and a diminished reach in the MIRI band.

\section{Conclusions}
In this work, for the first time we constrain the lifetime of decaying DM using measurements from the James Webb Space Telescope (JWST).  
JWST is sensitive to wavelengths in the range 0.6 $\micron$ to 28.3 $\micron$, corresponding to DM masses between 0.1$\eV$ and 4.1$\eV$ for a two-photon decay signal.
% DM candidates with masses in the electron volt range could decay into photon(s) with a wavelength in the optical and infrared band. This is of particular interest for JWST, which is sensitive to wavelengths in the range (0.6,28.3) $\mu$m , corresponding to DM masses between 0.1 eV and 4.1 eV for a two-photon decay signal.
We make use of public blank sky observations, originally collected for the purpose of sky subtraction.

Among the four 
instruments 
%different detectors
aboard JWST, we consider here NIRSpec IFU measurements, which are well-suited for blank-sky DM searches. This work focuses on 
sky spectra collected in tandem with 
observations of the galaxy GN-z11.  
% With available blank sky data from the Near-Infrared Spectrograph aboard JWST, 
We find no significant evidence of DM line emission in this data and we considerably   
% With this data, we significantly 
improve the bounds 
on decaying DM in the range 0.8$\eV$ to 2.5$\eV$ compared to the current 
bounds %state-of-the-art 
in the literature.
We exclude a lifetime of $\tau < 2.4 \cdot 10^{26}$ s for a DM mass of 1$\eV$, and up to a maximal constraint of $\tau < 6.7\cdot 10^{26}$ s at $\mdm = 0.863 \eV$.
Among the numerous DM candidates in this mass range, axion-like particles (ALPs) are one of the most compelling. 
% In Fig. \ref{fig:limits} we translated the limits on the lifetime for a generic decaying DM particle into the axion parameter space, excluding axion-photon coupling $g_{a \gamma \gamma}$ of order $10^{11}$ s for masses of 1 eV.  
For ALP DM, we exclude photon couplings $g_{a\gamma\gamma} > 2.4 \cdot 10^{-11} \GeV^{-1}$ for $m_a = 1\,\eV$, down to $g_{a\gamma\gamma} > 9.4 \cdot 10^{-12} \GeV^{-1}$ for $m_a = 2.15 \eV$.

This work confirms the competitiveness of infrared astronomy to search for and constrain compelling DM candidates.
In particular, our blank sky approach will allow this DM search to be updated and strengthened as JWST continues to operate, regardless of the specific targets chosen for future observations.  
We predict that by using all currently collected data we will be able to exclude lifetimes 
%between $4.7 \cdot 10^{27} \sec$ and $1.9 \cdot 10^{28} \sec$, 
up to $4.3 \cdot 10^{27}$ s, corresponding to ALP-photon couplings 
%of $1.8 \cdot 10^{-12} \GeV^{-1}$ to $3.2 \cdot 10^{-12} \GeV^{-1}$. 
down to $3.7 \cdot 10^{-12} \GeV^{-1}$, across the mass window of NIRSpec, $0.5 \eV$ to $4 \eV$.
We leave this analysis for a future work.
With 15 years of data taking, we predict that this may be improved to lifetimes
%between $1.3 \cdot 10^{28} \sec$ and $5.1 \cdot 10^{28} \sec$, 
up to $10^{28}$ s and 
and ALP-photon couplings 
%of $1.1 \cdot 10^{-12} \GeV^{-1}$ to $1.9 \cdot 10^{-12} \GeV^{-1}$. 
down to $ 2 \cdot 10^{-12} \GeV^{-1}$.

\label{sec:conclusion}

\acknowledgments
We would like to thank Jorge Martin Camalich, Anirudh Chiti, Shany Danieli, Alexander Ji, Kristen McQuinn, Marcia Rieke, Jan Scholtz, Jorge Terol, Brian Welch for valuable conversations, and in particular Albert Stebbins for crucial guidance. 
This work makes use of observations made with the NASA/ESA/CSA James Webb Space Telescope. The data were obtained from the Mikulski Archive for Space Telescopes at the Space Telescope Science Institute, which is operated by the Association of Universities for Research in Astronomy, Inc., under NASA contract NAS 5-03127 for JWST. These observations are associated with program \#4426.
% from here: https://archive.stsci.edu/publishing/mission-acknowledgements
This work made use of Astropy \citep{astropy:2013, astropy:2018, astropy:2022}, Astroquery \citep{2019AJ....157...98G}, matplotlib \citep{Hunter:2007}, NumPy \citep{harris2020array}, and SciPy \citep{Virtanen_2020}.
This material is based upon work supported by the U.S. Department of Energy, Office of Science, National Quantum Information Science Research Centers, Superconducting Quantum Materials and Systems Center (SQMS) under contract number DE-AC02-07CH11359.
We acknowledge support from {\sl Fermi Research Alliance, LLC} under Contract No. DE-AC02-07CH11359 with the U.S. Department of Energy, Office of High Energy Physics.

\bibliographystyle{JHEP}
\bibliography{ref}

\providecommand{\href}[2]{#2}\begingroup\raggedright\begin{thebibliography}{10}

\bibitem{Boddy:2022knd}
K.K.~Boddy et~al., \emph{{Snowmass2021 theory frontier white paper: Astrophysical and cosmological probes of dark matter}}, \href{https://doi.org/10.1016/j.jheap.2022.06.005}{\emph{JHEAp} {\bfseries 35} (2022) 112} [\href{https://arxiv.org/abs/2203.06380}{{\ttfamily 2203.06380}}].

\bibitem{Krnjaic:2023odw}
G.~Krnjaic and E.~Pinetti, \emph{{Probing The Longest Dark Matter Lifetimes with the Line Emission Mapper}},  \href{https://arxiv.org/abs/2307.00041}{{\ttfamily 2307.00041}}.

\bibitem{Regis:2020fhw}
M.~Regis, M.~Taoso, D.~Vaz, J.~Brinchmann, S.L.~Zoutendijk, N.F.~Bouch\'e et~al., \emph{{Searching for light in the darkness: Bounds on ALP dark matter with the optical MUSE-faint survey}}, \href{https://doi.org/10.1016/j.physletb.2021.136075}{\emph{Phys. Lett. B} {\bfseries 814} (2021) 136075} [\href{https://arxiv.org/abs/2009.01310}{{\ttfamily 2009.01310}}].

\bibitem{Caputo:2020msf}
A.~Caputo, A.~Vittino, N.~Fornengo, M.~Regis and M.~Taoso, \emph{{Searching for axion-like particle decay in the near-infrared background: an updated analysis}}, \href{https://doi.org/10.1088/1475-7516/2021/05/046}{\emph{JCAP} {\bfseries 05} (2021) 046} [\href{https://arxiv.org/abs/2012.09179}{{\ttfamily 2012.09179}}].

\bibitem{Todarello:2023hdk}
E.~Todarello, M.~Regis, J.~Reynoso-Cordova, M.~Taoso, D.~Vaz, J.~Brinchmann et~al., \emph{{Robust bounds on ALP dark matter from dwarf spheroidal galaxies in the optical MUSE-Faint survey}},  \href{https://arxiv.org/abs/2307.07403}{{\ttfamily 2307.07403}}.

\bibitem{Grin:2006aw}
D.~Grin, G.~Covone, J.-P.~Kneib, M.~Kamionkowski, A.~Blain and E.~Jullo, \emph{{A Telescope Search for Decaying Relic Axions}}, \href{https://doi.org/10.1103/PhysRevD.75.105018}{\emph{Phys. Rev. D} {\bfseries 75} (2007) 105018} [\href{https://arxiv.org/abs/astro-ph/0611502}{{\ttfamily astro-ph/0611502}}].

\bibitem{Carenza:2023qxh}
P.~Carenza, G.~Lucente and E.~Vitagliano, \emph{{Probing the blue axion with cosmic optical background anisotropies}}, \href{https://doi.org/10.1103/PhysRevD.107.083032}{\emph{Phys. Rev. D} {\bfseries 107} (2023) 083032} [\href{https://arxiv.org/abs/2301.06560}{{\ttfamily 2301.06560}}].

\bibitem{Nakayama:2022jza}
K.~Nakayama and W.~Yin, \emph{{Anisotropic cosmic optical background bound for decaying dark matter in light of the LORRI anomaly}}, \href{https://doi.org/10.1103/PhysRevD.106.103505}{\emph{Phys. Rev. D} {\bfseries 106} (2022) 103505} [\href{https://arxiv.org/abs/2205.01079}{{\ttfamily 2205.01079}}].

\bibitem{Bessho:2022yyu}
T.~Bessho, Y.~Ikeda and W.~Yin, \emph{{Indirect detection of eV dark matter via infrared spectroscopy}}, \href{https://doi.org/10.1103/PhysRevD.106.095025}{\emph{Phys. Rev. D} {\bfseries 106} (2022) 095025} [\href{https://arxiv.org/abs/2208.05975}{{\ttfamily 2208.05975}}].

\bibitem{Adams:2022pbo}
C.B.~Adams et~al., \emph{{Axion Dark Matter}},  in \emph{{Snowmass 2021}}, 3, 2022 [\href{https://arxiv.org/abs/2203.14923}{{\ttfamily 2203.14923}}].

\bibitem{PhysRevD.16.1791}
R.D.~Peccei and H.R.~Quinn, \emph{Constraints imposed by $\mathrm{CP}$ conservation in the presence of pseudoparticles}, \href{https://doi.org/10.1103/PhysRevD.16.1791}{\emph{Phys. Rev. D} {\bfseries 16} (1977) 1791}.

\bibitem{PhysRevLett.38.1440}
R.D.~Peccei and H.R.~Quinn, \emph{$\mathrm{CP}$ conservation in the presence of pseudoparticles}, \href{https://doi.org/10.1103/PhysRevLett.38.1440}{\emph{Phys. Rev. Lett.} {\bfseries 38} (1977) 1440}.

\bibitem{PhysRevLett.40.279}
F.~Wilczek, \emph{Problem of strong $p$ and $t$ invariance in the presence of instantons}, \href{https://doi.org/10.1103/PhysRevLett.40.279}{\emph{Phys. Rev. Lett.} {\bfseries 40} (1978) 279}.

\bibitem{PhysRevD.11.3583}
S.~Weinberg, \emph{The u(1) problem}, \href{https://doi.org/10.1103/PhysRevD.11.3583}{\emph{Phys. Rev. D} {\bfseries 11} (1975) 3583}.

\bibitem{Cadamuro:2011fd}
D.~Cadamuro and J.~Redondo, \emph{{Cosmological bounds on pseudo Nambu-Goldstone bosons}}, \href{https://doi.org/10.1088/1475-7516/2012/02/032}{\emph{JCAP} {\bfseries 02} (2012) 032} [\href{https://arxiv.org/abs/1110.2895}{{\ttfamily 1110.2895}}].

\bibitem{Navarro:1995iw}
J.F.~Navarro, C.S.~Frenk and S.D.M.~White, \emph{{The Structure of cold dark matter halos}}, \href{https://doi.org/10.1086/177173}{\emph{Astrophys. J.} {\bfseries 462} (1996) 563} [\href{https://arxiv.org/abs/astro-ph/9508025}{{\ttfamily astro-ph/9508025}}].

\bibitem{Cirelli:2010xx}
M.~Cirelli, G.~Corcella, A.~Hektor, G.~Hutsi, M.~Kadastik, P.~Panci et~al., \emph{{PPPC 4 DM ID: A Poor Particle Physicist Cookbook for Dark Matter Indirect Detection}}, \href{https://doi.org/10.1088/1475-7516/2012/10/E01}{\emph{JCAP} {\bfseries 03} (2011) 051} [\href{https://arxiv.org/abs/1012.4515}{{\ttfamily 1012.4515}}].

\bibitem{Cirelli:2020bpc}
M.~Cirelli, N.~Fornengo, B.J.~Kavanagh and E.~Pinetti, \emph{{Integral X-ray constraints on sub-GeV Dark Matter}}, \href{https://doi.org/10.1103/PhysRevD.103.063022}{\emph{Phys. Rev. D} {\bfseries 103} (2021) 063022} [\href{https://arxiv.org/abs/2007.11493}{{\ttfamily 2007.11493}}].

\bibitem{NIRSpec-Dispersers-Filters}
``Nirspec dispersers and filters.'' \url{https://jwst-docs.stsci.edu/jwst-near-infrared-spectrograph/nirspec-instrumentation/nirspec-dispersers-and-filters}, 2017.

\bibitem{Evans:2018bqy}
N.W.~Evans, C.A.J.~O'Hare and C.~McCabe, \emph{{Refinement of the standard halo model for dark matter searches in light of the Gaia Sausage}}, \href{https://doi.org/10.1103/PhysRevD.99.023012}{\emph{Phys. Rev. D} {\bfseries 99} (2019) 023012} [\href{https://arxiv.org/abs/1810.11468}{{\ttfamily 1810.11468}}].

\bibitem{Drukier:1986tm}
A.K.~Drukier, K.~Freese and D.N.~Spergel, \emph{{Detecting Cold Dark Matter Candidates}}, \href{https://doi.org/10.1103/PhysRevD.33.3495}{\emph{Phys. Rev. D} {\bfseries 33} (1986) 3495}.

\bibitem{2022A&A...661A..80J}
P.~{Jakobsen}, P.~{Ferruit}, C.~{Alves de Oliveira}, S.~{Arribas}, G.~{Bagnasco}, R.~{Barho} et~al., \emph{{The Near-Infrared Spectrograph (NIRSpec) on the James Webb Space Telescope. I. Overview of the instrument and its capabilities}}, \href{https://doi.org/10.1051/0004-6361/202142663}{\emph{Astron. Astrophys.} {\bfseries 661} (2022) A80} [\href{https://arxiv.org/abs/2202.03305}{{\ttfamily 2202.03305}}].

\bibitem{Gardner:2006ky}
J.P.~Gardner et~al., \emph{{The James Webb Space Telescope}}, \href{https://doi.org/10.1007/s11214-006-8315-7}{\emph{Space Sci. Rev.} {\bfseries 123} (2006) 485} [\href{https://arxiv.org/abs/astro-ph/0606175}{{\ttfamily astro-ph/0606175}}].

\bibitem{JWST-background}
``Jwst background model.'' \url{https://jwst-docs.stsci.edu/jwst-general-support/jwst-background-model}, year=2017, note = {Accessed: Accessed: 2023-08-07}.

\bibitem{Rauscher:2007ta}
B.J.~Rauscher et~al., \emph{{Detectors for the James Webb Space Telescope Near-Infrared Spectrograph. 1. Readout Mode, Noise Model, and Calibration Considerations}}, \href{https://doi.org/10.1086/520887}{\emph{Publ. Astron. Soc. Pac.} {\bfseries 119} (2007) 768} [\href{https://arxiv.org/abs/0706.2344}{{\ttfamily 0706.2344}}].

\bibitem{2022A&A...661A..82B}
T.~{B{\"o}ker}, S.~{Arribas}, N.~{L{\"u}tzgendorf}, C.~{Alves de Oliveira}, T.L.~{Beck}, S.~{Birkmann} et~al., \emph{{The Near-Infrared Spectrograph (NIRSpec) on the James Webb Space Telescope. III. Integral-field spectroscopy}}, \href{https://doi.org/10.1051/0004-6361/202142589}{\emph{Astron. Astrophys.} {\bfseries 661} (2022) A82} [\href{https://arxiv.org/abs/2202.03308}{{\ttfamily 2202.03308}}].

\bibitem{Maiolino:2023wwm}
R.~Maiolino et~al., \emph{{JWST-JADES. Possible Population III signatures at z=10.6 in the halo of GN-z11}},  \href{https://arxiv.org/abs/2306.00953}{{\ttfamily 2306.00953}}.

\bibitem{2023arXiv230609142S}
J.~{Scholtz}, C.~{Witten}, N.~{Laporte}, H.~{Ubler}, M.~{Perna}, R.~{Maiolino} et~al., \emph{{GN-z11: The environment of an AGN at $z=$10.603}}, \href{https://doi.org/10.48550/arXiv.2306.09142}{\emph{arXiv e-prints} (2023) arXiv:2306.09142} [\href{https://arxiv.org/abs/2306.09142}{{\ttfamily 2306.09142}}].

\bibitem{Schlegel:1997yv}
D.J.~Schlegel, D.P.~Finkbeiner and M.~Davis, \emph{{Maps of dust IR emission for use in estimation of reddening and CMBR foregrounds}}, \href{https://doi.org/10.1086/305772}{\emph{Astrophys. J.} {\bfseries 500} (1998) 525} [\href{https://arxiv.org/abs/astro-ph/9710327}{{\ttfamily astro-ph/9710327}}].

\bibitem{extinction}
``Galactic dust reddening and extinction.'' \url{https://irsa.ipac.caltech.edu/applications/DUST/}.

\bibitem{2012ApJ...754...53K}
J.E.~{Krick}, W.J.~{Glaccum}, S.J.~{Carey}, P.J.~{Lowrance}, J.A.~{Surace}, J.G.~{Ingalls} et~al., \emph{{A Spitzer/IRAC Measure of the Zodiacal Light}}, \href{https://doi.org/10.1088/0004-637X/754/1/53}{\emph{Astrophys. J.} {\bfseries 754} (2012) 53} [\href{https://arxiv.org/abs/1205.4047}{{\ttfamily 1205.4047}}].

\bibitem{2023PASP..135d8002R}
J.R.~{Rigby}, P.A.~{Lightsey}, M.~{Garc{\'\i}a Mar{\'\i}n}, C.W.~{Bowers}, E.C.~{Smith}, A.~{Glasse} et~al., \emph{{How Dark the Sky: The JWST Backgrounds}}, \href{https://doi.org/10.1088/1538-3873/acbcf4}{\emph{Publications of the Astronomical Society of the Pacific} {\bfseries 135} (2023) 048002} [\href{https://arxiv.org/abs/2211.09890}{{\ttfamily 2211.09890}}].

\bibitem{CAST:2007jps}
{\scshape CAST} collaboration, \emph{{An Improved limit on the axion-photon coupling from the CAST experiment}}, \href{https://doi.org/10.1088/1475-7516/2007/04/010}{\emph{JCAP} {\bfseries 04} (2007) 010} [\href{https://arxiv.org/abs/hep-ex/0702006}{{\ttfamily hep-ex/0702006}}].

\bibitem{CAST:2017uph}
{\scshape CAST} collaboration, \emph{{New CAST Limit on the Axion-Photon Interaction}}, \href{https://doi.org/10.1038/nphys4109}{\emph{Nature Phys.} {\bfseries 13} (2017) 584} [\href{https://arxiv.org/abs/1705.02290}{{\ttfamily 1705.02290}}].

\bibitem{Ayala:2014pea}
A.~Ayala, I.~Dom\'\i{}nguez, M.~Giannotti, A.~Mirizzi and O.~Straniero, \emph{{Revisiting the bound on axion-photon coupling from Globular Clusters}}, \href{https://doi.org/10.1103/PhysRevLett.113.191302}{\emph{Phys. Rev. Lett.} {\bfseries 113} (2014) 191302} [\href{https://arxiv.org/abs/1406.6053}{{\ttfamily 1406.6053}}].

\bibitem{Dolan:2022kul}
M.J.~Dolan, F.J.~Hiskens and R.R.~Volkas, \emph{{Advancing globular cluster constraints on the axion-photon coupling}}, \href{https://doi.org/10.1088/1475-7516/2022/10/096}{\emph{JCAP} {\bfseries 10} (2022) 096} [\href{https://arxiv.org/abs/2207.03102}{{\ttfamily 2207.03102}}].

\bibitem{2023ApJ...943...95S}
C.~{Severino} and I.~{Lopes}, \emph{{Asteroseismology: Looking for Axions in the Red Supergiant Star Alpha Ori}}, \href{https://doi.org/10.3847/1538-4357/aca897}{\emph{Astrophys. J.} {\bfseries 943} (2023) 95} [\href{https://arxiv.org/abs/2212.01890}{{\ttfamily 2212.01890}}].

\bibitem{AxionLimits}
C.~O'Hare, ``cajohare/axionlimits: Axionlimits.'' \url{https://cajohare.github.io/AxionLimits/}, July, 2020.
\newblock 10.5281/zenodo.3932430.

\bibitem{Caputo:2021eaa}
A.~Caputo, C.A.J.~O'Hare, A.J.~Millar and E.~Vitagliano, \emph{{Dark photon limits: a cookbook}},  \href{https://arxiv.org/abs/2105.04565}{{\ttfamily 2105.04565}}.

\bibitem{Cowan:2010js}
G.~Cowan, K.~Cranmer, E.~Gross and O.~Vitells, \emph{{Asymptotic formulae for likelihood-based tests of new physics}}, \href{https://doi.org/10.1140/epjc/s10052-011-1554-0}{\emph{Eur. Phys. J. C} {\bfseries 71} (2011) 1554} [\href{https://arxiv.org/abs/1007.1727}{{\ttfamily 1007.1727}}].

\bibitem{Roach:2022lgo}
B.M.~Roach, S.~Rossland, K.C.Y.~Ng, K.~Perez, J.F.~Beacom, B.W.~Grefenstette et~al., \emph{{Long-exposure NuSTAR constraints on decaying dark matter in the Galactic halo}}, \href{https://doi.org/10.1103/PhysRevD.107.023009}{\emph{Phys. Rev. D} {\bfseries 107} (2023) 023009} [\href{https://arxiv.org/abs/2207.04572}{{\ttfamily 2207.04572}}].

\bibitem{2015ApJ...808..124S}
J.~{Strader}, A.K.~{Dupree} and G.H.~{Smith}, \emph{{The 10830 {\r{A}} Helium Line Among Evolved Stars in the Globular Cluster M4}}, \href{https://doi.org/10.1088/0004-637X/808/2/124}{\emph{Astrophys. J.} {\bfseries 808} (2015) 124} [\href{https://arxiv.org/abs/1506.06307}{{\ttfamily 1506.06307}}].

\bibitem{2018ApJ...855L..11O}
A.~{Oklop{\v{c}}i{\'c}} and C.M.~{Hirata}, \emph{{A New Window into Escaping Exoplanet Atmospheres: 10830 {\r{A}} Line of Helium}}, \href{https://doi.org/10.3847/2041-8213/aaada9}{\emph{Astrophys. J. Letters} {\bfseries 855} (2018) L11} [\href{https://arxiv.org/abs/1711.05269}{{\ttfamily 1711.05269}}].

\bibitem{Cowan:2011an}
G.~Cowan, K.~Cranmer, E.~Gross and O.~Vitells, \emph{{Power-Constrained Limits}},  \href{https://arxiv.org/abs/1105.3166}{{\ttfamily 1105.3166}}.

\bibitem{astropy:2013}
{Astropy Collaboration}, T.P.~{Robitaille}, E.J.~{Tollerud}, P.~{Greenfield}, M.~{Droettboom}, E.~{Bray} et~al., \emph{{Astropy: A community Python package for astronomy}}, \href{https://doi.org/10.1051/0004-6361/201322068}{\emph{Astron. Astrophys.} {\bfseries 558} (2013) A33} [\href{https://arxiv.org/abs/1307.6212}{{\ttfamily 1307.6212}}].

\bibitem{astropy:2018}
{Astropy Collaboration}, A.M.~{Price-Whelan}, B.M.~{Sip{\H{o}}cz}, H.M.~{G{\"u}nther}, P.L.~{Lim}, S.M.~{Crawford} et~al., \emph{{The Astropy Project: Building an Open-science Project and Status of the v2.0 Core Package}}, \href{https://doi.org/10.3847/1538-3881/aabc4f}{\emph{The Astronomical Journal} {\bfseries 156} (2018) 123} [\href{https://arxiv.org/abs/1801.02634}{{\ttfamily 1801.02634}}].

\bibitem{astropy:2022}
{Astropy Collaboration}, A.M.~{Price-Whelan}, P.L.~{Lim}, N.~{Earl}, N.~{Starkman}, L.~{Bradley} et~al., \emph{{The Astropy Project: Sustaining and Growing a Community-oriented Open-source Project and the Latest Major Release (v5.0) of the Core Package}}, \href{https://doi.org/10.3847/1538-4357/ac7c74}{\emph{apj} {\bfseries 935} (2022) 167} [\href{https://arxiv.org/abs/2206.14220}{{\ttfamily 2206.14220}}].

\bibitem{2019AJ....157...98G}
A.~{Ginsburg}, B.M.~{Sip{\H o}cz}, C.E.~{Brasseur}, P.S.~{Cowperthwaite}, M.W.~{Craig}, C.~{Deil} et~al., \emph{{astroquery: An Astronomical Web-querying Package in Python}}, \href{https://doi.org/10.3847/1538-3881/aafc33}{\emph{The Astronomical Journal} {\bfseries 157} (2019) 98} [\href{https://arxiv.org/abs/1901.04520}{{\ttfamily 1901.04520}}].

\bibitem{Hunter:2007}
J.D.~Hunter, \emph{Matplotlib: A 2d graphics environment}, {\emph{Computing In Science \& Engineering} {\bfseries 9} (2007) 90}.

\bibitem{harris2020array}
C.R.~Harris, K.J.~Millman, S.J.~van~der Walt, R.~Gommers, P.~Virtanen, D.~Cournapeau et~al., \emph{Array programming with {NumPy}}, \href{https://doi.org/10.1038/s41586-020-2649-2}{\emph{Nature} {\bfseries 585} (2020) 357}.

\bibitem{Virtanen_2020}
P.~{Virtanen}, R.~{Gommers}, T.E.~{Oliphant}, M.~{Haberland}, T.~{Reddy}, D.~{Cournapeau} et~al., \emph{{SciPy 1.0: Fundamental Algorithms for Scientific Computing in Python}}, \href{https://doi.org/https://doi.org/10.1038/s41592-019-0686-2}{\emph{Nature Methods} {\bfseries 17} (2020) 261}.

\end{thebibliography}\endgroup

\end{document}